# A multiball read-out for the spherical proportional counter

**A. Giganon, I. Giomataris, M. Gros, I. Katsioulas*, X. F. Navick, G. Tsiledakis**

*IRFU, CEA, Université Paris-Saclay, F-91191 Gif-sur-Yvette, France*

**I. Savvidis**

*Aristotle University of Thessaloniki, Thessaloniki, Greece*

**A. Dastgheibi-Fard**

*LSM, CNRS/IN2P3, Université Grenoble-Alpes, Modane, France*

**A. Brossard**

*Department of Physics, Engineering Physics & Astronomy, Queen's University, Kingston, Ontario K7L 3N6, Canada*

*Corresponding author
  E-mail: ioannis.katsioulas@cea.fr

ABSTRACT: We present a novel concept of proportional gas amplification for the read-out of the spherical proportional counter. The standard single-ball read-out presents limitations for large diameter spherical detectors and high pressure operations. We have developed a multi-ball read-out system which consists of several balls sitting at a fixed distance from the center of the spherical vessel. Such a module can tune the volume electric field at the desired value and can also provide detector segmentation with individual ball read-out. In the latter case the large volume of the vessel becomes a spherical time projection chamber with 3D capabilities.

# Contents

1. Introduction

The development of low-background, low-energy threshold detectors is presently a challenge in low-energy neutrino physics [1] and Dark Matter (DM) search [2].

The detection of low energy neutrinos (below a few tens of MeV) via coherent nuclear scattering is an experimental challenge because the energy transfer to the nucleus is very small, less than 1 keV for a typical reactor neutrino at energies around 3 MeV. Several suggestions have been proposed to measure this process from several neutrino sources (Sun, Supernovae, Earth, reactors, stopped-pion beams and spallation sources) using different detection techniques [3-7].

The search for DM in the form of hypothetical Weakly Interacting Massive Particles (WIMPs) is under intense development and relies on the detection of low energy (keV scale) recoils produced by their elastic nuclear interaction with detector nuclei.
The direct detection of DM particles with mass below a few GeV scale is an experimental challenge related to the detection of "soft" nuclear recoils and the required low-energy detector threshold. The advantage of using detectors able to reach sub-keV energy threshold in order to improve detector sensitivity and capacity to detect light mass WIMPs has been pointed out by many authors[8-10]. To meet the requirements for such rare event search, a novel gaseous spherical detector has been developed with the ability to operate at pressure up to 10 bar, with a volume of a few $m^3$ of gas with various light targets such as H, He and Ne nuclei [11]. A DM detector made of such light element targets is more sensitive in detecting light WIMPs in the GeV and sub-GeV range, than current world-leading experiments made of Xe and Ge which are generally focusing on larger masses. The NEWS-G project [12-13] is dedicated to the direct search for very-low mass WIMPs, in the 0.1 to 10 GeV mass range.

The detector [14-17] is a gaseous Spherical Proportional Counter (SPC), based on the radial geometry, which has been recently developed in Saclay by I. Giomataris and collaborators. It uses a small anode ball as an amplification structure located at the center of a large spherical gaseous vessel at ground potential. A significant advantage of this structure is the use of a single electronic channel to read-out a large volume. This information alone still allows the determination of the radial coordinate of the interaction point through the measurement of the time dispersion of the measured charge pulse. Such information is of paramount importance for localization in depth and background rejection by applying fiducial cuts [17-18]. Such an approach simplifies the construction and reduces the cost of the project. Large gains were obtained providing low energy threshold in the sub-keV region. The radial electric field is distorted because of the rod that sustains the anode ball. To restore the radial field a simple field corrector was used providing a good energy resolution over the whole volume of the spherical vessel [16].

Another challenge is the weakness of the volume electric field when a large spherical detector is required. In the following sections we present details of this particular problem along with an elegant solution which can be used for any size of the spherical vessel.

## 2. Electric field challenge

In the spherical proportional counter, the electric field $E$ depends on the anode radius $r_1$, the cathode radius $r_2$, the anode voltage $V_0$ and the radial distance $r$:

$$E(r) = \frac{V_0}{r^2} \frac{1}{1/r_1 - 1/r_2} \qquad (1)$$

Since $r_1 \ll r_2$ the electric field formula can be approximated by

$$E(r) = \frac{V_0}{r^2} r_1 \qquad (2)$$

The required anode ball size is about 10 mm in diameter for low pressure up to a few hundred mbar, several mm for pressure up to a few bar and of the order of 1 mm for much higher pressure.

The operation of the SPC in high pressure over a few bar is limited by the contradiction between the need for high absolute gain and reduced drift times (the time required for the primary ionization charges to reach the anode). Both the drift velocity and the absolute gain $M$ depend on the reduced electric field ($E/P$) where $P$ is the pressure in the vessel:

$$lnM = \int_{E(r_1)}^{E(r_2)} a(E/P) \frac{dr}{dE} dE \qquad (3)$$

$$v_{drift} = \mu \frac{E}{P} \qquad (4)$$

From equation (2) we observe that at a large distance (in order of 1 m) the electric field becomes very weak for applied voltage up to about 10 kV resulting in very low values of reduced electric field and therefore to very high drift times. The solution is to increase the radius of the anode ball as much as possible. This is contradicted by the need for high gain values. The achievement of the same gain for higher pressure requires the increment of the high voltage applied and the decrement of the anode ball radius. Large increment of the high voltage is limited by the creation of discharges between the anode ball and the ground or the field corrector. Such limitations may become a concern for large detectors. A similar behavior occurs if the gas pressure is increased, requiring a reduction of the ball size in order to reach comparable *E/P*.

The solution to this problem is to replace the single-ball by a multi-ball read out system, called ACHINOS. According to the choice of parameters - such as the distance from the center of the detector and the diameter of the balls - it is then possible to tune the electric field while maintaining the gain high. An additional advantage of this device is that it allows segmentation of the detector.

## 3. ACHINOS multi-ball sensor: design and simulation

The ACHINOS sensor consists of a set of anode balls uniformly distributed around a central sphere at an equal distance from the center of the detector as shown on figure 1. The anode balls are supported by insulated wires (at present, the insulation of these wires is made of Kapton) applying the high voltage ($HV_1$) on the anodes. These wires have a constant length and are fixed perpendicularly to the surface of the central sphere in such a way that all the anode balls are on a virtual sphere larger than the central sphere.

The central sphere is made of either highly resistive material or a total insulator covered by a conducting surface, thus making it possible to apply $HV_2$ on its surface (bias electrode). The value of $HV_2$ is chosen to optimize the electric field configuration and improve the energy resolution of the detector. This ball is held in the center of the detector by a hollow metallic rod through which the different wires pass. In order to preserve the regular polyhedron symmetry, this metallic rod is replacing one of the anode spheres. Up until now few ACHINOS sensors have been built with five, eleven and thirty three balls (see section 4). We have nevertheless limited our simulation to the eleven-ball-case.

After testing different materials for the production of the central sphere (including silicon, which can also be used) we have found that bakelite is best suited because of its resistivity is in the range of $10^{12}$ Ohm·cm.

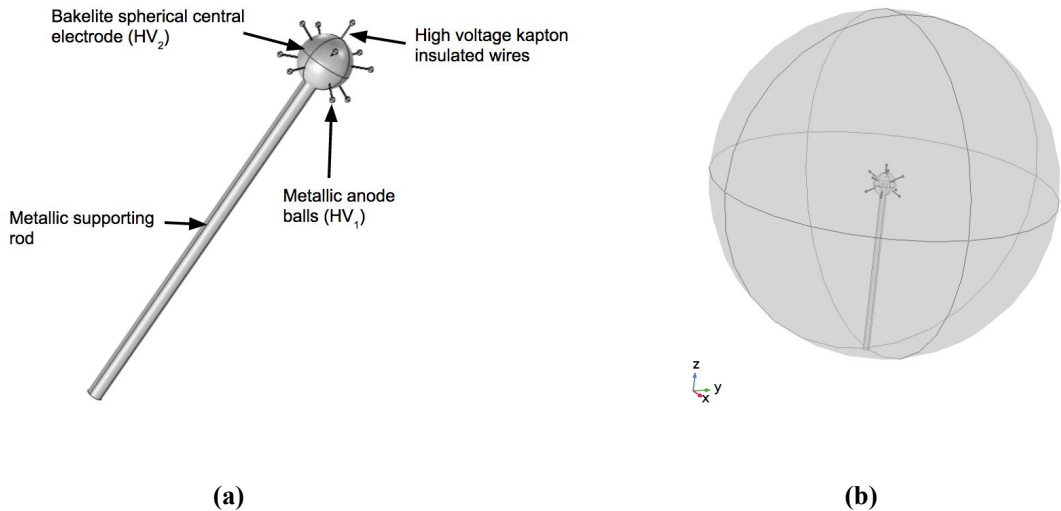

**(a)**  **(b)**

**Figure 1.** (a) The design of the multi-ball module ACHINOS. The balls are placed at an equal distance from the center of the resistive ball, with a regular distribution among a canonical polyhedron (here in the center of the faces of a canonical dodecahedron as an example). (b) The simulated geometry of the detector with the ACHINOS sensor mounted. The central axis of the rod is aligned with the *Z*-axis and the center of the bakelite ball is the origin of the coordinate system.

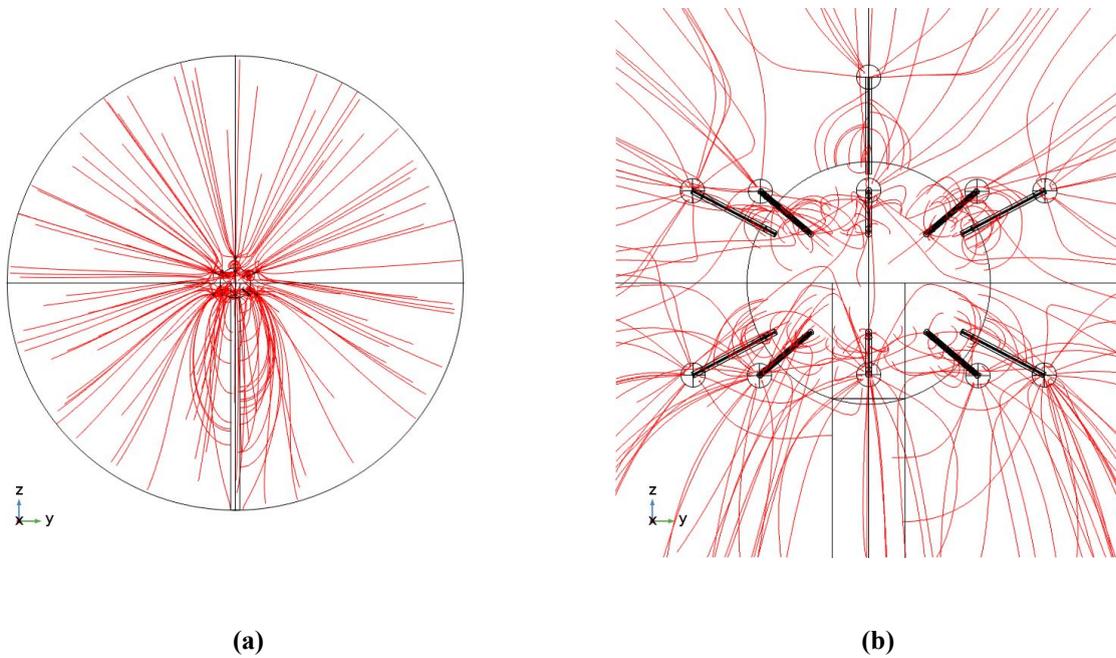

**(a)**  **(b)**

**Figure 2.** Streamlines of the electric field are displayed in three dimensions for $HV_1$ = 2000 V and $HV_2$ = 0 V for (a) the whole geometry of the detector and (b) focusing on the anode balls and the central electrode.

The principle of the design was studied extensively using COMSOL [20]. The model used in these simulations was an ACHINOS sensor with eleven 2-mm in diameter metallic balls (ACH-11), each placed at 7 mm from the surface of a Bakelite ball of 20-mm in diameter. A reference 2000 V electric potential was applied on the anode balls, whereas the potential applied on the central electrode was zero. The mechanical design of the ACH-11 sensor is presented in figure 1a along with the geometrical elements used for the simulation. The central axis of the rod is aligned with the *Z*-axis of the cartesian coordinate system which has its origin in the center of the Bakelite spherical ball.

Results of the simulations display the principle behind ACHINOS. Electrons liberated by ionization in the volume of the detector will drift under the influence of the electric field of the eleven balls equivalent to the case of single ball approximately 36 mm in diameter (figure 3). Their drift will lead them close to an anode ball, where the electric field is increasing strongly and the charge multiplication will take place (The avalanche process begins a few hundred μm from the anode ball surface).

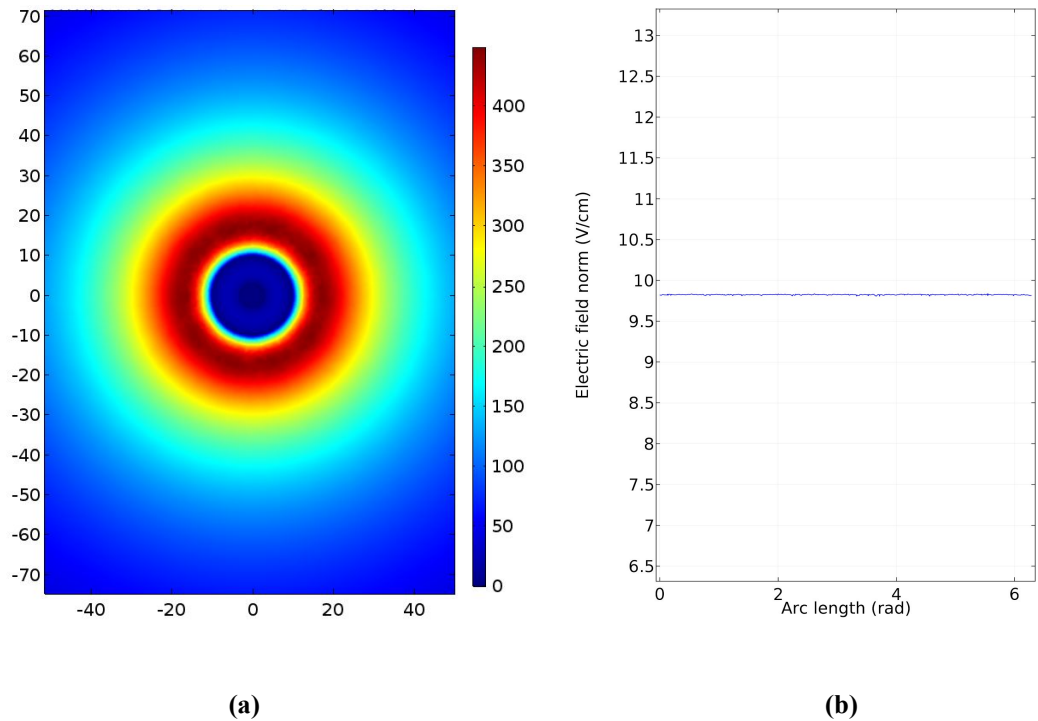

**(a)**          **(b)**

**Figure 3** (a) The contours of the electric potential at the *X-Y* plane for *Z=0* (center of the SPC). The effect of the ACHINOS module is visible, equipotential lines are formed similar to that of an anode with a diameter of 36 mm. (b) The plot corresponds to the value of the electric field magnitude as a function of the azimuthal angle ($\varphi$) from 0 to $2\pi$ at a radius of 10 cm in the *X-Y* plane (*Z=0*).

As mentioned in the introduction, the electric field around the anode deviates from the ideal case due to the thin wire. In the case of ACHINOS the effect is limited because only the 'north' hemispheres of the anode balls (the hemisphere of each anode ball opposite to the wire) contribute to the avalanche process where the field is very homogenous as displayed in figure 4.

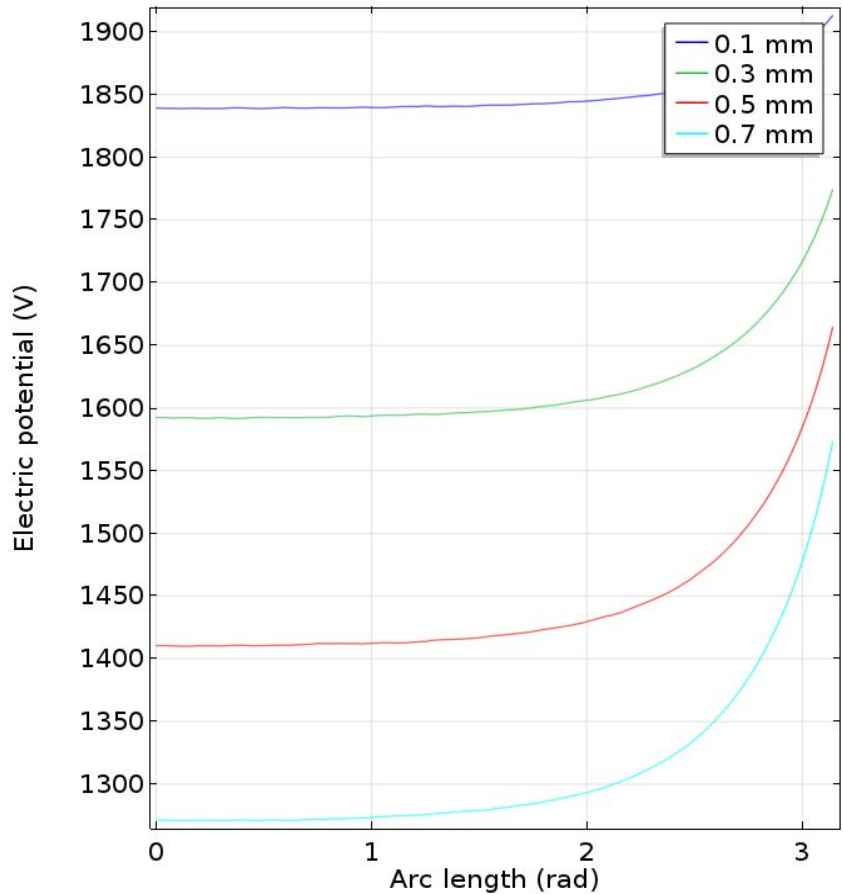

**Figure 4.** The electric potential around an anode ball at a distance of 0.1 mm, 0.3 mm, 0.5 mm, 0.7 mm from the surface of the anode versus the polar angle (the coordinate system has its center at the center of the ball the *Y* axis is parallel to the wire).

Electric field lines are presented in figure 2 for eleven 2 mm in diameter anode spheres uniformly distributed on a 36-mm-in-diameter virtual sphere surrounding a 20 mm-in diameter central spherical bias electrode for $HV_1 = 2000$ V and $HV_2 = 0$ V.

The difference between a single ball sensor and the ACHINOS set at the same potential is clearly demonstrated in figure 5. The plot displays the magnitude of the electric field starting from the surface of the anode placed parallel to the *Z*-axis ("north") and compares it with the electric field of a single ball sensor with its anode in the center of the detector, placed on the surface of a Bakelite ball of the same radius as in the ACHINOS sensor. In both cases an electric potential of 2000 V is applied. The result is that the electric field close to the surface of the shell of the detector (300-mm diameter) is approximately 8 times higher in the case of the ACHINOS sensor than in the case of the single ball sensor.

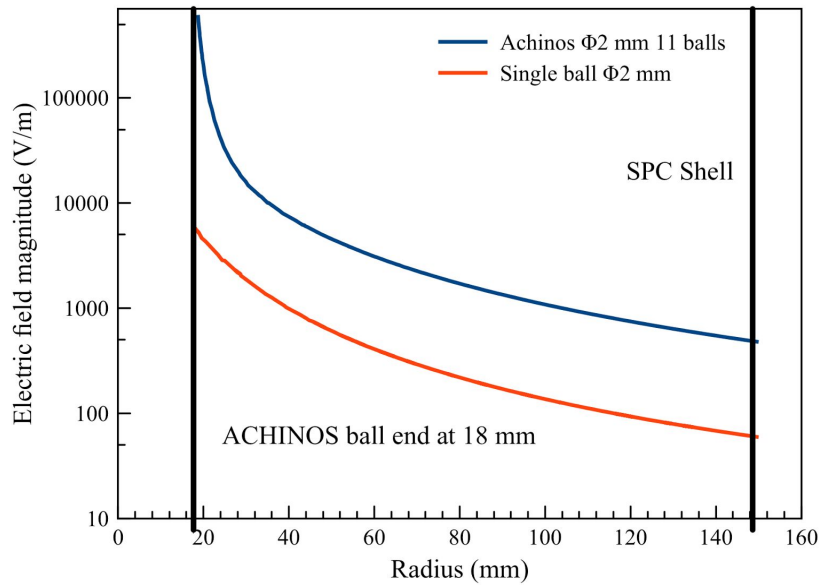

**Figure 5** Electric field magnitude versus the radius inside the detector in the case of a single 2-mm in diameter anode ball placed in the center of the 300-mm diameter detector and in the case of the ACHINOS sensor made of eleven balls distributed on a 36-mm diameter sphere placed in the center of the detector. The electric field in the far region from the center is ~*8* times higher for a detector equipped with an ACHINOS sensor than a single ball sensor placed at the same bias.

An additional advantage of the use of ACHINOS sensor is that by developing dedicated electronics each ball could be read-out separately. In this case the SPC would become a simple spherical Time Projection Chamber (TPC), allowing the segmentation of the detector.

## 4. Production of ACHINOS structures and measurements in test chamber

In this section we will describe the realization of the new multi-ball structure ACHINOS and give results of measurements obtained in the above laboratories.

Following the ACHINOS design described in section 3, three prototypes have been produced with 5, 11 and 33 metallic balls respectively. Metallic balls (1 to 2 mm in diameter) have been fixed and uniformly distributed around a bakelite ball. The diameter of the bakelite ball was 25 mm for the first two prototypes and 30 mm for the third one. The balls are connected with the high voltage power supply, through a small cable having a 150-μm thick wire and surrounded by a 200-μm thick insulating material generally Kapton. The cable can sustain up to 10 kV. A bias voltage can be set to the bakelite ball, which is used to improve the electric field uniformity. This is important in order to have the same amplification for any direction in the sphere. The distance between the bakelite ball and the metallic balls is 7 mm. This distance has been chosen to avoid sparks and to have a good homogeneity in the electric field. The test chamber is a 30-cm in diameter stainless steel sphere whose wall is 3 mm thick. Figure 6 presents a cut of the 5-ball ACHINOS and figure 7 a pictures of the first three prototypes.

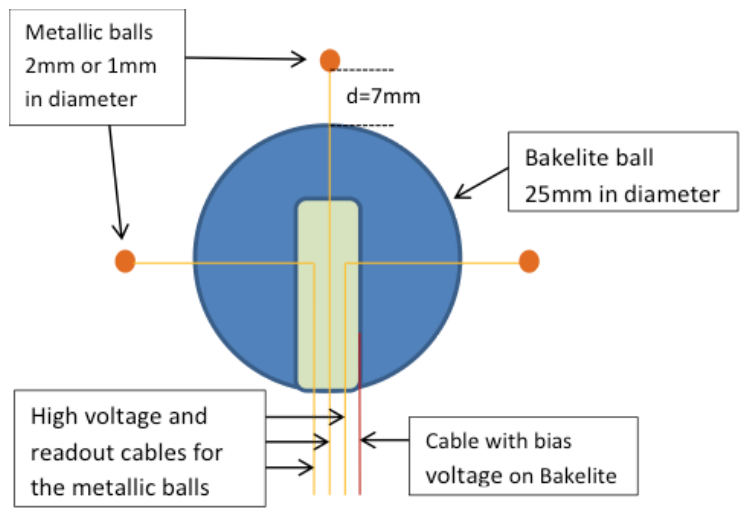

**Figure 6.** A cut of the 5-ball ACHINOS.

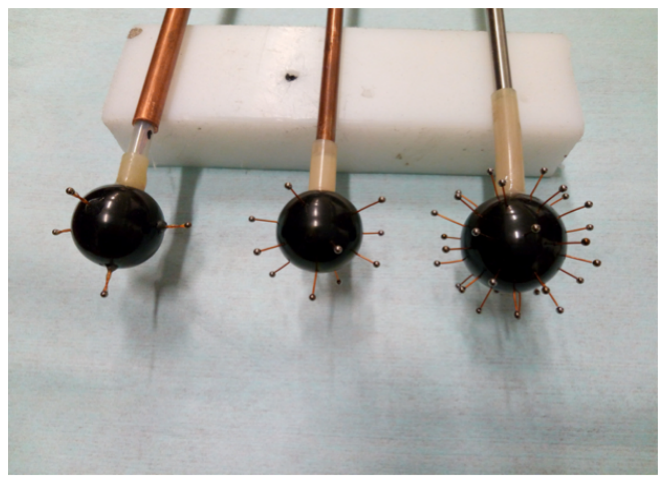

**Figure 7.** The three ACHINOS prototypes with 5, 11 and 33 metallic balls of 2 mm in diameter.

The ACHINOS structure, sustained by a metallic tube (4 mm inner and 6 mm outer diameter), was connected to the outer metallic sphere, both set at ground potential. The spherical shell of the detector made of 3-mm thick stainless steel is able to sustain high pressures (up to 10 bar).

There are two possible configurations for the reading out of the signals. The first possibility is to connect all the cables together to have one signal for each detected particle and the second possibility is to have independent signals from each metallic ball, which can lead to a TPC detection system. For the present tests we chose to apply the first configuration.

The volume was pumped by a primary pump followed by a turbomolecular pump, reaching a level of vacuum of about $10^{-6}$ mbar. Then He:Ar:$CH_4$ (80:11:9) gas was introduced up to 650 mbar.

Testing of the ACHINOS has been done using $^{55}$Fe source (5.9 keV X-rays). This radioactive source has been placed inside the sphere in such a way that it can move along the inner spherical surface by a magnet placed on the outer surface. With this arrangement it would be possible to scan the entire sphere and control electric field and detector gain uniformity. The pulse height distribution of the 5.9 keV X-ray line of the source as resolved by the SPC equipped with an 11-ball ACHINOS sensor is presented at figure 8. The 5.9 keV of the $^{55}$Fe source is measured with a resolution of 12.4 % which is limited by the signal induced with a different gain at some anode balls. This gain difference is displayed as a departure from the Gaussian form of the pulse height peak and the appearance of peaks with a lower mean value in the same spectrum. A full study of gain and uniformity will be presented in a forthcoming article.

We are limiting here the study to see the extension of rise time [18] of the pulse. This is correlated to the volume of electric field. As presented in figure 9, the full width of the rise time distribution is lower by a factor of ~8.7 for detectors equipped with an ACHINOS sensor in comparison with a single anode ball sensor (~1.1 μs and ~9.6 μs respectively). This reduction in the rise time dispersion is an indicator of the reduction in the drift time dispersion of the primary electrons and clearly displays the working principle of ACHINOS sensor.

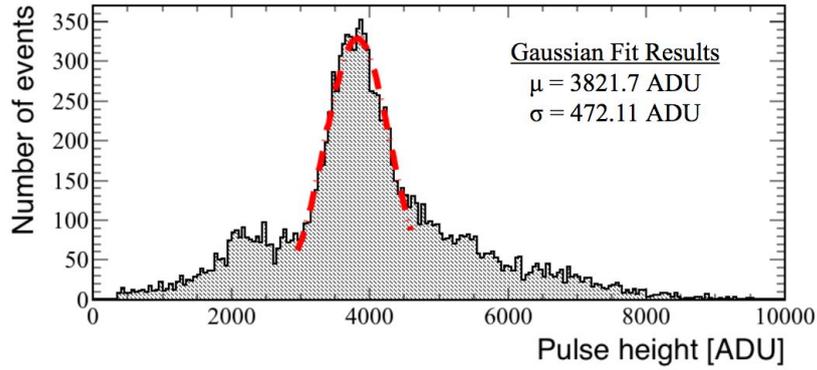

**Figure 8.** Pulse height distribution of the signal produced by the 5.9 keV X-ray line of an $^{55}$Fe source. The SPC was filled with a He:Ar:CH$_4$ (80:11:9) gas mixture at 640 mbar. The high voltage applied on the 2-mm in diameter anode ($HV_1$) was *2015 V* and on the central electrode ($HV_2$) was -200 V.

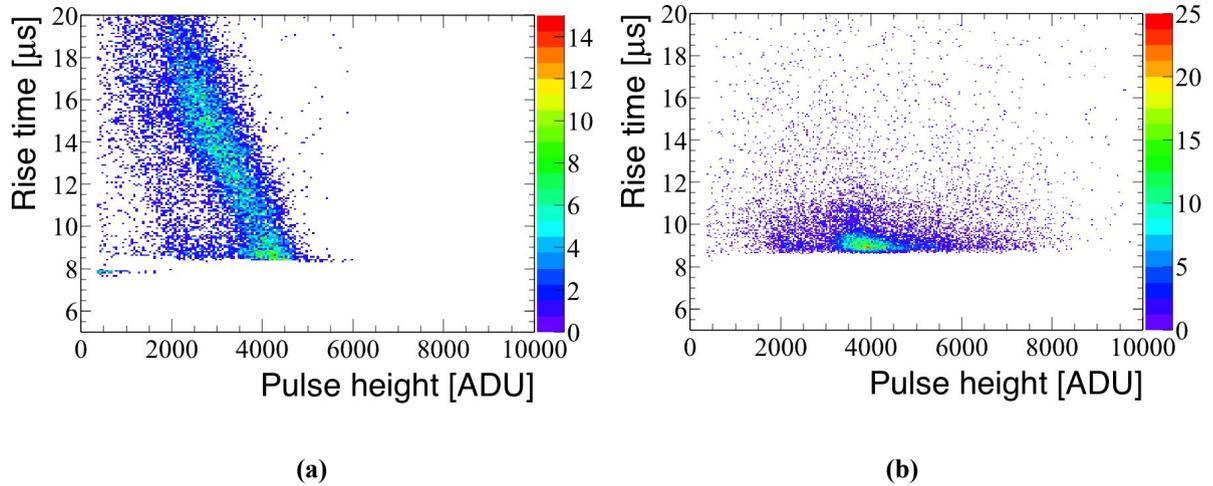

(a)  (b)

**Figure 9.** The pulse height versus the pulse rise time measured by a SPC filled with a He:Ar:CH$_4$ (80:11:9) gas mixture at 640 mbar in (a) the case of a single anode sensor equipped with a 2-mm anode ball, the high voltage applied on the anode ($HV_1$) was 2015 V  (b) the case of an 11-ball ACHINOS sensor with 2 mm anode balls, the high voltage applied on the anode ($HV_1$) was 2015 V and on the central electrode ($HV_2$) was -200 V. The rise time distribution of the single ball sensor has a full width of 9.6 μs whereas the one of ACHINOS has a full width of 1.1 μs.

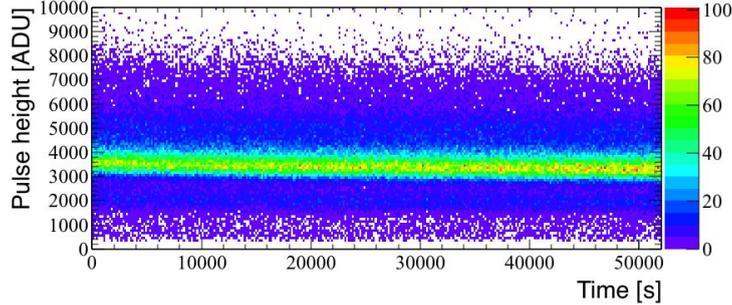

**Figure 10.** Pulse height evolution with time of the signal produced by a 5.9 keV $^{55}$Fe source. The SPC was filled with a He:Ar:CH$_4$ (80:11:9) gas mixture at 640 mbar. The high voltage applied on the 2-mm in diameter anode (HV$_1$) was 2015 V and on the central electrode ($HV_2$) was -200 V.

Long time acquisition have been taken over 50000 s to check the stability of the detector response, as seen in figure 10. A small drift of the gain has been observed but overall this result is satisfactory.

A full study of performances of detectors equipped with ACHINOS is underway to assess the uniformity of the detector response as well as the homogeneity of gain, and to optimize the biasing, with different virtual sphere radii and also changing the gas. Results will be presented in a forthcoming article.

## 5. Future developments and applications

Significant improvement in performances seems possible through the industrialization of the production of ACHINOS to obtain greater precision leading to a better uniformity of the electric field. A first approach is already under study, which includes the construction of the ACHINOS structure using 3D printing technology, as presented in figure 11. This structure will be tested in the near future.

A radio-pure selection of material is required to reduce the radioactive background for a DM search. We also plan to use ACHINOS in the detector developed for NEWS-SNOLAB whose diameter is 140 cm. The use of even smaller anode balls will be considered.
In addition, we will develop a dedicated ASIC to equip each individual anode ball with its own read-out. Such electronics will provide a 3D capability like in Time Projection Chambers.

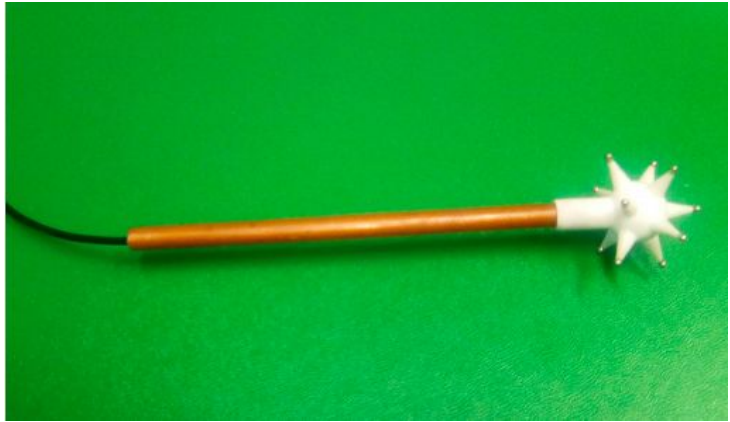

**Figure 11.** An 11-ball ACHINOS module already build using the 3D printing technology.

## 6.  Conclusions

We have developed a novel multi-ball read-out sensor that permits the tuning of the electric field in the volume of the spherical detector. The ACHINOS sensor will allow the detector to operate at very high pressure and very large detector volumes. An immediate application of this new design would improve charge collection efficiency of the NEWS-SNOLAB experiment.

The proposed spherical gaseous detector for the detection of very low mass particles of Dark Matter can also be used for other applications, but with a different size, constraints and operational conditions. The novel multi-ball sensor approach will satisfy the requirements of many projects which call for large target mass such as Supernova neutrino detection, neutrino-less double beta decay search with $^{136}$Xe pressurized gas and other similar applications. Such a large volume-mass detector system could be used to study low energy neutrino physics and detect the elastic neutrino-nucleus coherent scattering and could detect neutrinos from galactic Supernova explosions. A worldwide network consisting of several such simple and low-cost supernova detectors could be considered [7]. 

Large detector mass and high-pressure, 50 bar, Xenon spherical detectors will be competitive for future double beta decay experiments. A high pressure Xenon detector with good energy resolution could also be used for X-ray and Gamma ray spectroscopy as an alternative to conventional room temperature spectrometers (CdZnTe, LaBr3:Ce).

**Acknowledgments**

This work is funded by the French National Research Agency (ANR-15-CE31-0008).